\documentclass[runningheads]{llncs}
\usepackage[T1]{fontenc}
\usepackage{graphicx}
\usepackage{booktabs}
\usepackage{array}
\usepackage{amsmath}
\usepackage{hyperref}
\usepackage{tabularx}
\usepackage{biblatex}
\usepackage{fontawesome5}
\usepackage{soul}
\begin{document}
\title{Systematic Integration of Digital Twins and Constrained LLMs for Interpretable Cyber-Physical Anomaly Detection}
%
%
\author{Konstantinos E. Kampourakis\inst{1}\orcidID{0009-0000-8883-0735}\faIcon{envelope} \and
Vasileios Gkioulos \inst{1}\orcidID{0000-0001-7304-3835} \and
Sokratis Katsikas \inst{1}\orcidID{0000-0003-2966-9683}}
\authorrunning{Konstantinos E. Kampourakis}
%
\institute{Norwegian University of Science and Technology (NTNU), 2802 Gjøvik, Norway \email{\{konstantinos.kampourakis, vasileios.gkioulos, sokratis.katsikas\}@ntnu.no}}
\maketitle              
\begin{abstract}
Cyber attacks targeting Industrial Control Systems (ICS) have become increasingly sophisticated and hard to identify. Detecting such attacks requires integrating low-level behavioral cues with high-level semantic interpretation, a capability that traditional anomaly detectors lack. This paper presents a Digital Twin (DT)-driven hybrid detection approach that combines deterministic heuristics with systematic, constrained Large Language Model (LLM) reasoning to achieve real-time incident detection. The DT maintains a synchronized, feature-enriched representation of the Secure Water Treatment (SWaT) process, deriving behavioral descriptors. Heuristics identify characteristic signatures of spoofing, valve forcing, denial-of-service, and bias drift, while the LLM is invoked only when heuristics abstain. A constrained JSON schema and semantic plausibility filters ensure physically consistent LLM outputs, and a temporal smoothing layer stabilizes the final decision signal. Evaluation on four canonical SWaT attack scenarios shows that the proposed detector precisely localizes each attack interval with low time-to-detect and zero False Positives (FPs) in the evaluated benign region. Results are consistent across both a local LLaMA model and a cloud-based GPT model, demonstrating the robustness of the constrained hybrid architecture. The findings highlight the potential of DT-guided LLM reasoning as a reliable and interpretable approach to ICS anomaly detection.

\end{abstract}
\section{Introduction}
\label{S:Intro}

Industrial Control Systems (ICS) operate critical infrastructure such as water treatment, power distribution, and oil and gas pipelines. Increasing connectivity has expanded the attack surface, enabling cyber–physical disruptions that directly impact process state and safety. Detecting these attacks in real-time is challenging, since physical process deviations can resemble normal operational variance, control actions are highly coupled (meaning a legitimate change in one variable requires corresponding changes in many others, easily obscuring malicious input), and anomaly detectors rarely explain why a window is suspicious. This problem motivates more advanced detection systems that combine process awareness, temporal stability, and constrained semantic reasoning.

Digital Twins (DTs) offer a promising solution to this problem. They maintain a synchronized virtual representation of physical processes to support monitoring and diagnostics. Nevertheless, most existing approaches rely on physics-based simulation or narrowly tailored anomaly detection models that lack semantic interpretability and robust decision-making~\cite{kampourakis2025digital}. Moreover, Large Language Models (LLMs) offer capabilities for interpreting process behavior semantically and explaining potential consequences. Yet, unconstrained LLMs introduce risks of hallucinations, physically implausible conclusions, and unstable decision boundaries~\cite{huang2025survey}. Recent work has shown the importance of constraining LLM output with structured domain knowledge, highlighting a path forward, but a systematic integration of LLM reasoning within a DT framework for ICS security has not yet been demonstrated.

This work addresses this gap by introducing a DT-driven hybrid detector that systematically combines deterministic heuristics with structured LLM reasoning to achieve real-time cyber–physical incident detection. The DT maintains a continuously synchronized representation of the Secure Water Treatment (SWaT) process, derives behavioral features such as slopes, variances, ranges, and actuator toggling patterns, and invokes LLM-based analysis only when lightweight heuristics abstain from producing a decision. The LLM is constrained through a strict JSON schema and additional semantic filters that ensure the physical plausibility of predicted tactics and attack paths. A temporal smoothing layer further stabilizes the final decision signal, reducing short-lived fluctuations while preserving responsiveness.

Our evaluation demonstrates that the proposed detector achieves near-perfect localization of four canonical SWaT attack scenarios, while maintaining low false positive (FP) rates across extended benign periods. The system detects all attacks with consistent Time-To-Detect (TTD) characteristics and exhibits stable behavior across both a local LLaMA-3.1 model and a cloud-based GPT-4.1-mini model. In contrast, a standard Isolation Forest (IF) baseline fails to detect several scenarios and produces inconsistent alarms.

The main contributions of this work are summarized as follows:

\begin{itemize}
    \item A DT–driven detection architecture that maintains a synchronized, window-based behavioral replica of an industrial process for cyber–physical monitoring.
    \item A process-aware heuristic layer that captures ICS Indicators of Compromise (IoC) and selectively gates LLM invocation to reduce FPs.
    \item A constrained LLM reasoning module integrated into the DT workflow to provide interpretable attack classification under strict schema and plausibility checks.
    \item An evaluation on the SWaT dataset demonstrating precise attack localization, low TTD, and stable behavior across both local and cloud-based LLM backends.
\end{itemize}

The remainder of this paper is organized as follows. The next section surveys related work on LLM-based detection. Section~\ref{S:DTA} presents the DT architecture used in our system. Section~\ref{S:Methodology} describes the hybrid detection methodology. Section~\ref{S:Experiments} outlines the experimental setup, and Section~\ref{S:Results} reports the detection results. Section~\ref{S:Dis} provides a technical discussion of the findings. Finally, Section~\ref{S:Conclusion} concludes the paper.

\section{Related work}
\label{S:Related}

In this section, we elaborate on the related work on LLM detection in ICS, concentrating on recently published works. We explicitly focus on experimental publications rather than theoretical ones.

The authors in~\cite{simoni2025morse} propose a specialized Retrieval-Augmented Generation (RAG) pipeline designed specifically for cybersecurity question answering. The system integrates two cascaded and complementary RAG modules. That is, a fast structured RAG with seven parallel retrievers and a fallback unstructured RAG, enabling broad coverage over both structured sources and unstructured text. The authors evaluate MoRSE on 600 cybersecurity questions, each validated by domain experts, and show that it consistently outperforms GPT-4, Gemini, Mixtral, and HackerGPT in relevance, similarity, correctness, and accuracy. Particularly, MoRSE achieves 84\% accuracy on CVE questions, versus 34\% for GPT-4, and leads all models in Elo-based comparisons using GPT-4 as a judge.

The work in~\cite{hassanin2025pllm} introduces PLLM-CS, a transformer-based intrusion detection system (IDS) specifically targeting satellite communication networks. The authors address the lack of effective IDS solutions for satellite systems, which face high operational complexity, sparse connectivity, and increasing vulnerability to DoS/DDoS-style cyber threats. PLLM-CS adapts transformer self-attention to model sequential network dynamics by converting multivariate telemetry into token sequences (``sentencing''), enabling long-range contextual reasoning over satellite-like traffic patterns. The model is evaluated on UNSW-NB15 and TON\_IoT datasets, and achieves 100\% accuracy across all metrics, outperforming both machine learning and deep learning baselines.

In~\cite{gokcimen2025novel}, the authors propose a comprehensive multi-layered security architecture designed to reduce hallucinations, prevent prompt-injection attacks, and enhance privacy protections for secure and reliable LLM deployments. The system combines an injection-detection pipeline based on fine-tuned transformer classifiers, a Kernel module for semantic preprocessing and entity extraction, a VectorDB for secure vectorized retrieval, and a RAG layer to ensure context-bounded responses. A cross-LLM validation mechanism computes an eligibility score along dimensions such as toxicity, banned-topic violations, sensitivity, and sentiment. Experimental results across several LLM families show that the system consistently removes injected content, filters unsafe or manipulative prompts, and blocks semantically unsafe translations or harmful queries.

The contribution in~\cite{tirulo2025llm} proposes a generative-AI-based cybersecurity framework for electric vehicle cyber-physical systems, where traditional IDS approaches struggle to detect novel or zero-day attacks. The authors develop a fine-tuned LLM, based on a distilled GPT-2-like architecture, and integrate it into an EV monitoring pipeline that analyzes CAN bus traffic, sensor data, and charging-protocol messages. The system performs both anomaly detection and generative threat simulation, using prompt engineering to synthesize realistic zero-day attack patterns. Evaluation across real datasets and synthetic zero-day scenarios shows high accuracy (>98\%), low false-positive rates, and low inference latency suitable for real-time EV operation.

The authors in~\cite{du2024mad} propose a LLM–driven framework for detecting multi-stage cyber attacks by aggregating and correlating heterogeneous security alerts. Instead of relying on fixed correlation rules or predefined attack graphs, the system uses carefully engineered prompts, combining chain-of-thought reasoning, domain knowledge, and strict output formatting, to guide the LLM through two key tasks: (i) alert aggregation into coherent hyper-alerts, and (ii) alert correlation into sequential attack stages. A validation mechanism ensures that all alerts are assigned to exactly one hyper-alert and resolves any duplicates or misses via a systematic, iterative LLM refinement. The authors show that MAD-LLM outperforms classical multi-stage detection methods such as GAC and MAAC in precision, recall, and attack-stage recovery, even when critical stages have only one or two alerts. 

The study in~\cite{jin2024crimson} introduces a specialized framework for enhancing strategic cybersecurity reasoning in LLMs by mapping CVEs and CTI narratives to MITRE ATT\&CK techniques using a novel, structured, retrieval-aware training pipeline. The authors construct a large synthetic dataset by aligning CVE and CTI descriptions with ATT\&CK tactics, validated by human experts, and train models using their methods to integrate retrieved ATT\&CK knowledge directly into the reasoning process. The paper also introduces a domain-fine-tuned embedding model that improves ATT\&CK-technique retrieval accuracy. Experimental results demonstrate significant improvements in AST-based structural accuracy and a reduction in spurious ATT\&CK predictions compared with standard LLaMA-2, Mistral, and GPT-3.5 baselines.

\section{DT architecture}
\label{S:DTA}

The DT in our system acts as a continuously synchronized virtual replica of the SWaT process. In this work, the DT is explicitly implemented as a replay-based architecture, realized through a custom HTTP service that reconstructs synchronized plant state windows and derived behavioral features from recorded telemetry. 

At the core of the DT is the replay engine, which provides a coherent, time-indexed view of the system by exposing a $30$-second sliding window of telemetry through a lightweight HTTP interface. All time windows have synchronized values for every process variable of interest, enabling the virtual replica to evolve in parallel with the corresponding physical process. 

Building on the replay layer, the DT incorporates a behavioral extraction model that computes derivatives, slopes, variances, ranges, and actuator toggling statistics from each window. These descriptors form an interpretable representation of the system's short-term behavior and are engineered to highlight deviations from normal operation. They are used both by the heuristic detector and the LLM-based reasoning module. In effect, the DT does not simply store raw telemetry but instead maintains an enriched, process-aware view of plant dynamics.

The architecture further integrates a semantic reasoning layer built around the LLM. When heuristics abstain, the DT promotes the enriched, process-aware window features to the LLM, which interprets the observed behavior within an ICS threat schema. The reasoning engine assesses tactic–technique alignment, evaluates physically plausible attack paths, and proposes corresponding mitigations. The DT therefore supports 
higher-order semantic interpretation that goes beyond thresholding or statistical anomaly detection.

Finally, the DT includes a temporal decision layer, which mitigates short-term variability. The DT applies a smoothing policy across subsequent time windows such that the final decision and indicators reflect systematic deviations from normal rather than random noise. This renders dependable indicators of an incident while supporting time-sensitive decision making.

\begin{figure} [!ht]
    \centering
    \includegraphics[width=1\textwidth]{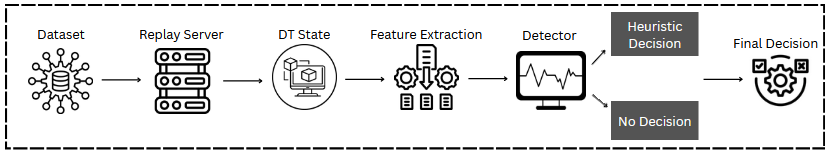}
    \caption{Overview of the DT-driven detection architecture.}
    \label{F:DT}
\end{figure}

Figure~\ref{F:DT} illustrates the overall DT architecture used in this work. The system begins with a replay server that streams a synchronized 30-second window of SWaT telemetry at each time step,  providing a virtualized DT state aligned with the physical process. This window is enriched with behavioral descriptors, such as slopes, variances, freeze ratios, and actuator toggles, that form the DT's interpretable feature-level representation of process dynamics. The hybrid detector then evaluates each window: deterministic process-aware heuristics first attempt to classify the behavior. If no clear pattern matches, the window is forwarded to the LLM reasoning module for schema-constrained semantic analysis. Both heuristic results and LLM outputs are merged through a temporal smoothing mechanism, which stabilizes fluctuations and produces the final, dependable attack decision signal.

\section{Methodology}
\label{S:Methodology}

In this section, we detail the design of our DT-driven hybrid detection pipeline. The system mirrors the live state of an industrial process, derives behavioral features, applies domain-specific heuristics, and invokes an LLM-based reasoning engine only when heuristics abstain from a decision. The methodology is structured around five interacting layers: (i) the replay and synchronization engine, (ii) the feature extraction module, (iii) the heuristic detector, (iv) the LLM reasoning agent, and (v) the temporal smoothing module.

\begin{itemize}
    \item \textit{\textbf{Replay server and temporal windowing:}} The DT keeps an up-to-date, synchronized view of the plant state while continuously replaying sensor and actuator values sourced from the SWaT dataset. A lightweight HTTP server facilitates access to a state endpoint, which returns a $30$-second sliding window at a given index. Each window contains aligned multivariate time series that correspond to sensors (e.g., LIT101, FIT101, AIT402) and actuators (e.g., MV101, P101), ensuring that the virtual process replica advances in synchrony with the underlying physical process. This mechanism acts as the DT's state-synchronization layer and is queried by the orchestrator at every step.

    \item \textit{\textbf{Behavioral feature and derivative extraction:}} For each window, the orchestrator computes a set of behavioral descriptors that capture short-term dynamics of the process. Specifically, these include, first-order slopes of key sensors, say $\Delta x_{i}(t)$, rolling standard deviations, amplitude ranges, flatness and freeze ratios, and actuator toggle counts. These features serve two purposes: (i) they enable efficient lightweight heuristics for common ICS attack patterns, and (ii) they provide a richer behavioral context for the LLM when heuristics abstain. The feature-extraction layer, therefore, forms the DT's behavioral model, a virtualization of system dynamics beyond raw telemetry.

     \item \textit{\textbf{Heuristic detector:}} Cyber–physical attacks often leave distinctive short-term behavioral signatures that can be identified without relying on high-level semantic reasoning. To exploit this, the DT incorporates a deterministic heuristic module that targets four characteristic patterns. In particular, spoofing is identified through sustained positive trends in LIT101 during pump or valve activity, captured by consistent slope–range behavior and adequate historical variance, which serves as an IoC for manipulated inflow. Valve Forcing manifests as rapid MV101 toggling accompanied by abnormal oscillatory patterns in FIT101 or LIT301. Bias Drift appears as low-variance but gradually diverging sensor readings, where the slope-to-standard-deviation ratio serves as a reliable IoC. Sensor-Freezing DoS is detected through multi-sensor flatness, suppressed variance, and elevated freeze ratios over extended windows. When any heuristic matches its associated IoC pattern with sufficient confidence, the detector produces a structured attack classification without invoking the LLM. 

     \item \textit{\textbf{LLM reasoning engine:}} If heuristics abstain, the orchestrator composes a structured prompt describing the window's telemetry, behavioral features, actuator states, and contextual history. The LLM is required to output a JSON object conforming to a strict schema containing: (i) a high-level ICS tactic and technique, (ii) a ranked list of hypothesized attack paths, (iii) suggested mitigations, and (iv) a confidence score. A schema validator enforces structural correctness and rejects insufficiently structured or hallucinated outputs. A post-filter then checks for semantic correctness, determining whether the stated tactics are physically possible given the state of the actuators observed during the test. 

     \item \textit{\textbf{Fusion and temporal smoothing:}} LLM predictions and heuristic detections are fused into a binary attack indicator. To reduce jitter and short-lived fluctuations, we apply a temporal smoothing rule that requires short-term majority agreement within a sliding decision window, typically set to $N$ time steps. This smoothing layer forms the DT's ``diagnostic stabilizer'', and ensures that detection decisions reflect persistent behavioral deviations rather than noise.

     \item \textit{\textbf{DT operational loop:}} The complete DT loop functions as follows: it first aligns the virtual plant state using the replay server; then derives behavioral features from each time window; applies heuristic rules to detect pattern-specific anomalies; triggers LLM-based reasoning only when no heuristic decision is made; then validates, filters, and smooths the resulting predictions; and finally emits structured incident metadata for downstream analysis. The architecture in Figure~\ref{F:DT} enables a real-time, semantically informed DT that can detect cyber-physical incidents with low false-positive rates and stable detection latencies across a wide range of attack scenarios.

\end{itemize}

In summary, Table~\ref{T:heuristics}, shows that the DT-driven hybrid detector applies deterministic heuristics, with each rule examining a distinct class of attack behavior commonly seen in SWaT. These heuristics are based on values that represent derived features in the application of slopes, variance metrics, actuation toggling patterns, and freeze ratios detected in the synchronized 30-second DT time frame. Specifically, sustained positive trends identified in LIT101 while flow components are active signal spoofing; rapid toggling of MV101 with oscillatory flow or level sensor activity captures valve forcing attacks; a low variance but steadily growing positive trend with a unique slope to standard deviation ratio is indicative of bias drift, and, the multi-channel flatness along with muted variance characterizes freezing DoS attacks due to a sensor being frozen. These rules enable rapid, interpretable, and physics-based attacks to be identified, without calling forth the LLM reasoning module, unless the behavior is ambiguous or falls outside the identified patterns.

\begin{table}[t]
\centering
\caption{Summary of heuristic rules used in the DT-driven detector.}
\label{T:heuristics}
\resizebox{\textwidth}{!}{%
\begin{tabular}{p{3cm} p{3.8cm} p{8cm}}
\toprule
\textbf{Attack Type} & \textbf{Signals} & \textbf{Heuristic Conditions} \\
\midrule

\textbf{Spoofing} 
& LIT101, FIT101, MV101, P101 
& Sustained positive slope in LIT101 under pump/valve activity; 
  slope--range consistency; sufficient historical variance (history~$\ge$~40). \\[6pt]

\textbf{Valve Forcing} 
& MV101, FIT101, LIT301 
& Rapid actuator toggling (MV101 toggles~$\ge 1$); 
  oscillatory behavior in FIT101 or LIT301 (range~$>0.03$); history~$\ge$~40. \\[6pt]

\textbf{Bias Drift} 
& LIT101 
& Gradually increasing or drifting trend with low variance; 
  slope~$>$~$\max(0.004, 1.2 \cdot \text{std})$; 
  standard deviation within $[0.005,0.06]$; 
  flatness ratio~$<0.90$; history~$\ge$~70. \\[6pt]

\textbf{Sensor-Freezing DoS} 
& LIT101, FIT101, AIT402 
& Multi-channel flatness; near-zero variance; high freeze ratios; 
  persistent flat behavior over long windows (history~$\ge$~80). \\

\bottomrule
\end{tabular}
}
\end{table}

\section{Experimental setup}
\label{S:Experiments}

The experimental evaluation assesses the effectiveness and stability of the proposed DT detector across four canonical cyber–physical attack scenarios in the SWaT testbed, which contains 449,920 time-stamped records of process telemetry and a binary attack label. Among these, 54,584 rows ($\approx12,1\%$) are labelled as attack and 395,336 rows ($\approx87,9\%$) as benign, reflecting a realistically imbalanced ICS operating regime. All experiments are conducted using the publicly available dataset, which contains normal operation interspersed with ground-truth attack intervals. The DT operates on a $30$-second sliding window and processes the dataset sequentially from index $0$ to $1400$, mirroring real-time conditions in which decisions must be made without access to future information. The 0–1,400 region contains the four injected scenarios considered in this study; later attack-labelled segments are out of scope for our detection coverage and are reserved for a 10k stability check. It is worth noting that the full 449k-sample SWaT dataset was not exhaustively replayed because the runtime bottleneck lies in evaluation time instrumentation, derivative logging, overlapping-window extraction, and schema validation, not in the detector itself. The hybrid model operates in a streaming, constant-cost manner and scales to arbitrarily long traces. Our extended 10k-window run confirms stable long-horizon behavior, indicating that scalability is not a limiting factor for deployment. 

\begin{figure}[t]
\centering
\includegraphics[width=0.9\linewidth]{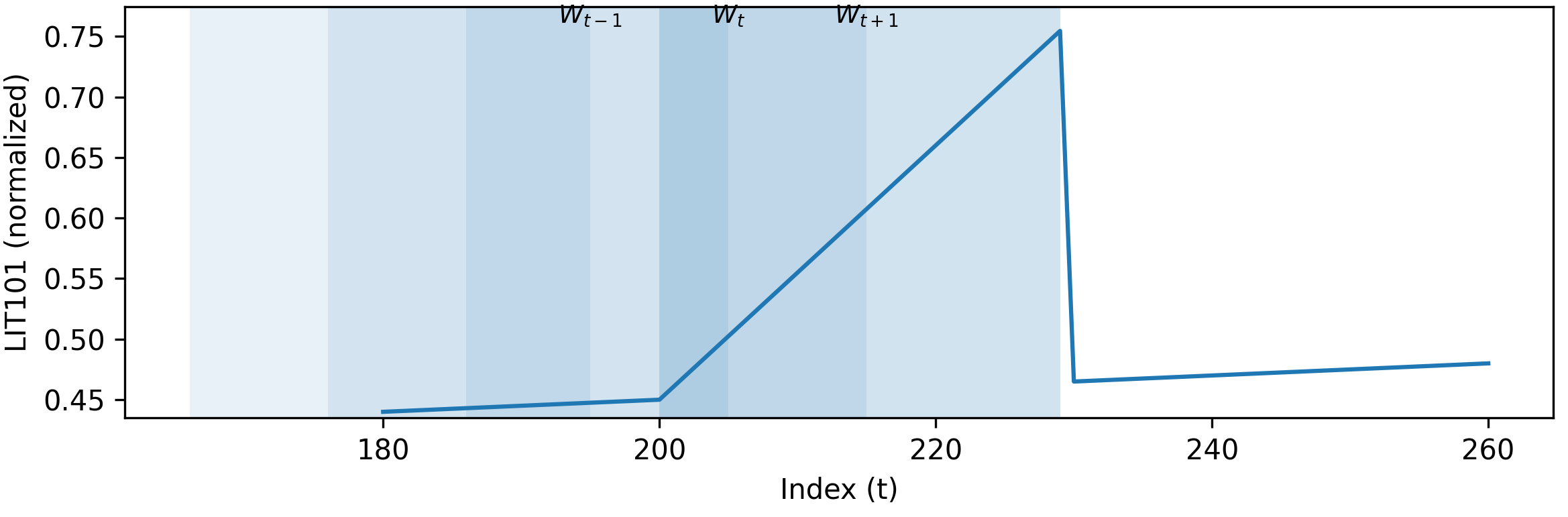}
\caption{Example of the DT replay windowing over SWaT-like telemetry.}
\label{F:windowing}
\end{figure}

Figure~\ref{F:windowing} shows the LIT101 level around the spoof interval (200--229), with the ground-truth attack region shaded. The three semi-transparent boxes illustrate consecutive 30-second DT windows ($W_{t-1}$, $W_t$, $W_{t+1}$) used for derivative computation, heuristic evaluation, and optional LLM reasoning.

The target attack scenarios include sensor spoofing (indices 200–229), valve-forcing (500–539), a sensor-freezing DoS attack (900–949), and a slow bias-drift manipulation of LIT101 (1200–1279). These scenarios span distinct physical mechanisms and behavioral signatures, making them appropriate for evaluating both the heuristic and LLM components of the detector. For each scenario, we measure (i) whether the system raises an alarm during the attack interval, (ii) the TTD, defined as the number of indices from attack onset to the first stabilized alarm, and (iii) the presence or absence of false alarms during benign periods.

Two LLM configurations are evaluated. The first uses a local deployment of LLaMA~3.1~8B running through a standard OpenAI-compatible inference server. The second uses the cloud-hosted GPT--4.1--mini model via the OpenAI API. Both models are invoked only when heuristics abstain. To ensure comparability, all system prompts, schema constraints, and semantic filters are identical across models, and the DT pipeline is executed end-to-end for both configurations.

For baseline comparison, we evaluate a standard Isolation Forest (IF) anomaly detector trained on the first 1,000 indices of benign data and applied to 15-second sliding windows. The IF baseline represents a commonly used unsupervised anomaly detection approach in ICS literature, but does not produce semantic explanations. IF is used as the baseline because it is a widely adopted, lightweight unsupervised anomaly detector in ICS security settings where labeled attack data and process models are unavailable. It relies purely on statistical deviation and does not capture temporal structure, actuator–sensor coupling, or semantic attack intent. As such, it provides a representative contrast to the proposed DT-driven hybrid detector, whose contribution lies in integrating process awareness, temporal stabilization, and constrained semantic reasoning rather than in competing with deep learning–based IDS approaches.

All detectors output structured JSON traces containing per-window predictions, confidence values, and smoothed binary decisions. Final evaluation is performed using a standardized post-processing script that aligns predictions with the ground-truth intervals via time-step indexing. The DT-driven hybrid detector is evaluated for both overall accuracy and temporal precision, with particular emphasis on whether alarms are confined to the correct attack windows without significant spillover into surrounding benign periods.

\section{Results}
\label{S:Results}

This section evaluates the performance of the proposed DT-driven hybrid detector across the four SWaT attack scenarios. We report detection success (binary), TTD, and FP behavior, and compare the two LLM configurations we used, LLaMA~3.1~8B and GPT--4.1--mini, to assess the stability of the hybrid reasoning layer. All metrics are computed using the standardized evaluation script described earlier.

\subsection*{Detection ranges and alignment with ground truth}

Table~\ref{T:LLM_results} shows that across all experiments, the DT-driven detector achieves precise localization of the four attack intervals. For both LLM backends, alarms occur strictly within the ground-truth windows for sensor spoofing (200--229), valve forcing (500--539), sensor-freezing DoS (900--949), and bias drift (1200--1279). No spillover is observed into the surrounding benign regions, indicating that the combination of heuristic gating and temporal smoothing effectively prevents unstable early firing.

We also ran the detector on the first 10k rows to check stability over long periods. In that run, the system raised alarms in only two continuous regions (4896–4902 and 7586–7597). We inspected those windows and confirmed they matched real abnormal behavior, not random or incorrect alarms. All other tested ranges that contained only normal process behavior produced no alarms.

Moreover, according to Table~\ref{T:IF_results}, the IF baseline does not demonstrate comparable precision. While IF successfully detects the DoS attack, it fails to raise consistent alarms during the spoofing, valve-forcing, and bias-drift intervals, and it produces scattered anomalies outside the ground-truth ranges.

\begin{table}[t]
\centering
\caption{Detection performance across the four SWaT attack scenarios for both LLM cases.}
\label{T:LLM_results}

\begin{tabular}{lccc}
\toprule
\textbf{Scenario} & \textbf{Ground Truth} & \textbf{Detected} & \textbf{TTD (windows)} \\
\midrule
Spoofing / Inflow   & 200--229   & 1 & 1  \\
Valve Forcing       & 500--539   & 1 & 1  \\
Sensor-Freezing DoS & 900--949   & 1 & 0  \\
Bias Drift          & 1200--1279 & 1 & 0  \\
\bottomrule
\end{tabular}
\end{table}

\begin{table}[t]
\centering
\caption{IF baseline results using 15-second windows.}
\label{T:IF_results}

\begin{tabularx}{\textwidth}{lcccc}
\toprule
\textbf{Scenario} & \textbf{Ground Truth} & \textbf{IF Detected} & \textbf{IF TTD} & \textbf{Notes} \\
\midrule
Spoofing / Inflow   & 200--229   & 0 & --- & No anomaly spike \\
Valve Forcing       & 500--539   & 0 & --- & Oscillations missed \\
Sensor-Freezing DoS & 900--949   & 1 & 0   & High freeze ratio  \\
Bias Drift          & 1200--1279 & 0 & --- & Slow drift undetected \\
\bottomrule
\end{tabularx}
\end{table}

\subsection*{Time-to-Detect}

In this case, TTD refers to the number of windows from scenario onset to the first hybrid alarm inside the matched interval.

The hybrid detector yields consistent TTD across scenarios and includes immediate detection of bias drift, reporting a TTD of 0. This indicates that the detector raised alarms on the first queried window falling fully inside the attack interval, without requiring additional history accumulation or semantic confirmation delays. DoS variance flattening likewise resulted in TTD 0, while spoofing and valve‐forcing each triggered within one window from onset.

Both LLM configurations produce nearly identical TTD values. Differences between LLaMA and GPT are within a single window across all scenarios, demonstrating that the LLM is used primarily for semantic confirmation rather than for primary detection.

\subsection*{False positive behavior}

A key requirement for deploying LLM-based reasoning in operational ICS settings is the minimization of hallucination-driven FPs. Table~\ref{T:false-positives} demonstrates that across the full 1400-index run, neither LLaMA nor GPT produces any false alarms outside the four attack intervals. This is attributable to (i) strict heuristic gating, which suppresses unnecessary LLM invocation, (ii) the constrained JSON schema enforced on LLM output, and (iii) the semantic plausibility filter that removes physically inconsistent threat reports.

By contrast, the IF baseline flags 273 benign windows as anomalous ($\approx$19,5\%), particularly in regions of transient sensor variability. These scattered detections do not align with any operational changes or ground-truth events.

\begin{table}[t]
\centering
\caption{FPs across the full 0--1400 index run.}
\label{T:false-positives}

\begin{tabular}{lcc}
\toprule
\textbf{Model} & \textbf{FPs} & \textbf{FP Regions} \\
\midrule
LLaMA-3.1-8B  & 0 & --- \\
GPT-4.1-mini  & 0 & --- \\
IF & 273 & scattered \\
\bottomrule
\end{tabular}
\end{table}

\subsection*{Comparison of LLaMA and GPT Backends}

The two LLM configurations deliver nearly identical detection behavior despite their architectural differences. Both models produce stable, schema-conforming outputs, and no hallucinated tactics or physically implausible attack paths are observed in the post-filtered predictions. 

GPT--4.1--mini shows marginally higher semantic confidence values in its threat reports, while LLaMA demonstrates slightly faster response times when deployed locally. However, these differences do not affect the combined decision signal due to the overriding influence of heuristics and temporal smoothing. This suggests that the DT-driven detector is robust to the choice of LLM backend, provided that schema constraints and gating mechanisms are enforced.

\subsection*{Summary}

Overall, the results demonstrate that the proposed DT-driven hybrid detector achieves real-time and high-precision detection across the four SWaT attack scenarios. The method consistently identifies the correct intervals, exhibits low TTD values, and maintains zero FPs in extended benign regions. The stable performance across both local and cloud LLM backends highlights the reliability of the constrained hybrid reasoning architecture.

Additionally, both LLaMA--3.1--8B and GPT--4.1--mini produce identical detection and TTD metrics after temporal smoothing; any model-level differences occur before fusion and are overridden by heuristic gating and temporal smoothing in the fused decision layer.

\section{Discussion}
\label{S:Dis}

The research results highlight several key characteristics of the proposed DT-based hybrid detection method, particularly in terms of stability, interpretability, and operational robustness.

A central outcome of the evaluation is that the hybrid reasoning approach proves to be dependable. The reliance on domain-specific heuristics to handle most detection tasks and only calling the LLM when strictly required means that the hybrid approach yields a more stable system than would be possible with unconstrained LLMs alone. Moreover, the ability to avoid spurious FPs over an extended period of benign activity indicates that, when tailored to a particular ICS process environment, heuristic rules significantly limit the likelihood of hallucination-induced alerts. The LLM supplies semantic reasoning and contextual validation; the LLM itself, however, is not the primary driver of anomaly detection. This division of responsibilities markedly shrinks the system’s attack surface while preserving the core benefits of high-level reasoning.

The DT architecture underpins long-term stability in several ways. It maintains a synchronized, feature-rich representation of process behavior over time, enabling both the heuristic layer and the LLM to operate in a controlled environment with a consistently defined temporal window. The DT’s behavioral model attenuates short-term fluctuations in process activity, directing the reasoning components toward significant patterns rather than transient noise. Moreover, the temporal smoothing mechanism further dampens the influence of short-lived deviations, ensuring that decisions are driven by data that is persistently or progressively manipulated. 

Additionally, the consistency with which LLaMA and GPT operate as model classes is another key observation: despite differences in their training processes, such as model size, training corpora, or deployment configuration, both behave similarly when used as LLM components within a DT framework. This suggests that the overall system behavior is driven more by the governing constraints, namely, heuristics, schema validators, and semantic filters, than by the specific architectural features that differentiate individual LLMs. Operationally, this benefits organizations, as they can flexibly choose between locally hosted and cloud-based LLMs based on factors like available resources, latency constraints, or cost, without significantly affecting detection fidelity.

The outcomes are promising, but they also underline the substantial complexity involved in identifying certain categories of cyber–physical attacks. One such example is bias drift, which manifests as a gradual, abnormal trend over time and therefore must be detected by examining temporal patterns and historical context. The hybrid detection process can capture these relationships through merging feature-based methods with semantic reasoning, yet its overall accuracy depends heavily on meticulous tuning of decision thresholds and temporal aggregation windows. 

In summary, the results validate the feasibility of integrating DT models, deterministic rule-based heuristics, and structured LLM reasoning into a unified methodology for real-time detection of cyber–physical incidents. Nevertheless, in this study, we tuned detection heuristics on SWaT dynamics and validated only for the 4 injected scenarios. Evaluation on other domains remains future work, and is not claimed in this work.

Despite the promising outcomes, we identify some limitations that guide future research. First, the detection heuristics are meticulously tuned to the SWaT process dynamics, specifically validated against only the four injected attack scenarios. This limits the generalizability of the current rule-set to diverse industrial domains or zero-day attacks. Second, identifying slow-manifesting attacks like bias drift requires precise calibration of temporal aggregation windows and decision thresholds to balance sensitivity against robustness. This highlights the critical trade-off between maximizing sensitivity to subtle manipulation and ensuring robustness against legitimate process fluctuations caused by load changes or operational routines. Third, the current DT relies on a behavioral mirror approach and does not incorporate physics-based modeling; integrating such models could further strengthen semantic correctness and expand the system's ability to detect novel, physically implausible states.

\section{Conclusions}
\label{S:Conclusion}

In this paper, we describe a DT-based hybrid detection approach that utilizes deterministic heuristics and constrained LLM reasoning to facilitate incident detection within ICS. The DT provides the means for the hybrid detection method to detect both dynamic and semantic characteristics of the process, through real-time synchronization with the SWaT process. The LLM is utilized only where the heuristics abstain from intervention, and a strict output schema is used with checks for semantic plausibility, which allows for the avoidance of hallucination-induced FPs that are present when using unconstrained LLMs.

Our evaluation demonstrates that the proposed method accurately localizes four canonical attack scenarios, sensor spoofing, valve forcing, DoS, and bias drift, with consistent TTD values and no spillover into benign regions. The detector performs reliably across both a locally hosted LLaMA model and a cloud-based GPT model, highlighting the robustness of the overall architecture and the relatively small influence of model-specific variability when strong structural constraints are applied. In contrast, a standard IF baseline fails to provide comparable temporal or semantic precision.

The findings suggest that DT representations enriched with behavioral features, combined with structured LLM reasoning, offer a promising direction for interpretable, resilient cyber–physical monitoring. Future work will explore the integration of physics-based modeling into the DT, extension to larger-scale industrial processes, online adaptation under concept drift, and the use of multi-agent LLM architectures for collaborative situational assessment. These directions aim to further advance the role of DTs as trustworthy components in next-generation ICS security frameworks.

\subsection*{Acknowledgments}
This work is supported by the Research Council of Norway through
the SFI Norwegian Centre for Cybersecurity in Critical Sectors (NORCICS) project no. 310105
%
%
%
%
\printbibliography
\end{document}